\documentstyle[aps,preprint]{revtex}
\begin{document}
\draft
\title{Order Parameters for Non-Abelian Gauge Theories}
\author{Hoi-Kwong Lo}
\address{
 School of Natural Sciences, Institute for Advanced Study, Olden Lane,\\
 Princeton, NJ 08540, U.S.A.
}
\date{\today}
\preprint{IASSNS-HEP-94/92}
\maketitle
\mediumtext
\begin{abstract}
Owing to subtle
issues concerning quantum fluctuations and gauge fixing, a
formulation of a general
procedure to specify the realization of non-Abelian gauge symmetry has evaded
all earlier attempts. In this Letter, we discuss these subtleties and
present two sets of order parameters for non-Abelian gauge theories.
Such operators directly probe the manifest low energy symmetry group
and are crucial for the study of the phase diagram of a non-Abelian
gauge theory.
\end{abstract}
\medskip
\pacs{PACS numbers:11.15.-q,03.65.Bz,03.80.+r}
%\narrowtext
A convenient way to investigate the phase diagram of a theory
is to construct a set of order
parameters which exhibit non-analytical behavior at phase boundaries.
For the case of Abelian gauge theories, this was done in a remarkable
paper by Preskill and Krauss \cite{PreKra}.
The key concept behind their construction
is the long ranged Aharonov-Bohm effect \cite{AhaBoh}: Free (as opposed
to screened or confined) charges inside a region can be detected by a
loop of cosmic string \cite{KraWil}. (Non-Abelian
vortices also Aharonov-Bohm scatter one another
\cite{WilcWu,Bucher,LoPres}.)
Generalization of this work to non-Abelian strings is far from
straightforward.
While much has been learnt about some subtle aspects of non-Abelian
gauge invariance\cite{LoPres,Alford,BuLoPr,BuLePr,Bavade},
a formulation of a general procedure to unambiguously specify
the realization of gauge symmetry has eluded all previous attempts
\cite{FreMar,AlfMar,qft}.

In this letter we present
two closely related sets of
order parameters for non-Abelian gauge theories
and discuss the subtle interplay of quantum fluctuations and
gauge fixing in the calibration and
measurement processes for the non-Abelian Aharonov-Bohm effect
\cite{LoPres,AlCoMa}.
One of these two sets of operators
was first constructed by
Alford {\em et al.\ \/} a couple of years ago \cite{qft},
but was immediately discarded
by them because in their original formulation
it was plagued by quantum fluctuations.
We will see below how this problem can be overcome.

The long
range Aharonov-Bohm effect---that the wave function of a charged particle
generally acquires a non-trivial phase upon covariant transport
around a cosmic string---has deep implications.
Since
no local operator can destroy a particle that
has an infinite range interaction
with another object, there are charge superselection
sectors. One of the consequences of this {\it quantum
mechanical\/} interaction is that black holes can carry quantum
numbers which are classically undetectable \cite{CoPrWi}.
One can study the phase diagram of a gauge theory by creating a
world sheet of cosmic string and investigating its Aharonov-Bohm
interactions with charged particles.
Such an experiment necessarily proceeds in two stages:
1) setting up a laboratory together with the calibration of
the string flux and the states of the charged particles and
2) subsequent interference experiment to measure the
non-Abelian Aharonov-Bohm phase.
An element, $g$, of the gauge group will be a manifest symmetry
at low energies if and only if
charged particles scattering
off a cosmic string with calibrated flux $g$
are able to recover the Aharonov-Bohm phase factor associated with
$g$.
In much of this Letter, we will consider the lattice
formulation of our operators \cite{PreKra,AlfMar,qft} and,
for simplicity, we will mainly discuss
the case where the gauge group is finite. Our results, however, can
be readily generalized to continuous gauge groups.

Let us consider the first stage in the Abelian case \cite{AlCoMa}.
First, we need an operator which inserts,
as a classical source, a string world sheet
of flux $a$. It was suggested in \cite{KraWil} that when a $U(1)$
gauge symmetry
is spontaneously broken into $Z_N$, the discrete $Z_N$ charge
$Q_{\Sigma^*}$ contained in a closed surface
$\Sigma^*$ can still be measured via
the Gauss law:
\begin{equation}
F(\Sigma^*)= \exp \left({2 \pi i \over N} Q_{\Sigma^*} \right)
=\exp  \left( {2 \pi i \over
Ne} \int_{\Sigma^*} E \cdot ds \right). \label{gauss}
\end{equation}
Now we turn to the operator which introduces classical charges into the
system, the Wilson loop
operator, $W^{\nu}(C)$ where $\nu$ is an irreducible representation of
the gauge group, $G$.
Therefore, one might naively expect
$F(\Sigma^*) W^{\nu}(C)$ to be the order parameter.
This is not quite correct because quantum mechanical fluctuations near
the surface
$\Sigma^* $ cause an area law decay of the {\em modulus\/} of
$F(\Sigma^*)$. Fortunately,
the {\em phase\/} of $F(\Sigma^*)$ remains unscreened and we can isolate
it by dividing out its vacuum expectation value\cite{PreKra} .
Similarly, quantum fluctuations also
lead to an exponential decay of the expectation value of $W(C)$.
Therefore, the true order parameter for Abelian gauge theories is
\cite{PreKra}
\begin{equation}
A^{\nu}(\Sigma^{*}, C) = {F(\Sigma^{*})
W^{\nu}(C) \over  \langle F(\Sigma^*) \rangle
\langle W^{\nu}(C)\rangle}  \label{abelian}
\end{equation}
where the limit that $\Sigma^{*}$ and $C$ are infinitely large and far apart
is taken. In a free charge phase, this gives
the expected Aharonov-Bohm phase.
We reserve the
discussion on the Higgs phase near the end of this
Letter\cite{PreKra,AlfMar,qft}.

When the gauge group is non-Abelian, the flux, $a$, of a string
is {\em not\/} a gauge invariant quantity. One can, however,
gauge fix by defining a basis of charged particles at some base point
far from the string loop and calibrate the string by scattering
those particles of known transformation properties from it
along a {\em particular\/} path.
Therefore, to specify the flux of a string, we have to
choose a base point
$x_0$, a basis
and a path that we use for the measurement.
In the lattice formulation, it is convenient to put a string world sheet
on a closed
surface $\Sigma^* $ on dual lattice. Let $\Sigma$ be the set of
plaquettes threaded by $\Sigma^*$.
Now, for each plaquette $P$
in $\Sigma$, we choose a path, $l_P$, that runs from the base point $x_0$ to a
corner of the
plaquette \cite{qft}. Calibration of the plaquette is done
along the path $l_P P l_P^{-1} $.
Suppose that the plaquette action is
\begin{equation}
   S^{(R)}_{gauge, P}= - \beta \chi^{(R)} (U_P) + c.c. \label{oldgauge}
\end{equation}
   where $R$ is some representation of the gauge group that defines the theory.
The insertion of $F_a(\Sigma^{*}, x_0, \{l_P\})$ modifies the action to
\begin{equation}
 S^{(R)}_{gauge, P} \to
 - \beta \chi^{(R)} (V_{l_P}a V_{l_P}^{-1} U_P) + c.c. \label{newgauge}
\end{equation}
where
\begin{equation}
V_{l_P} = \prod_{l \in l_P} U_l.  \label{path}
\end{equation}
This procedure can be easily generalized to insert {\em coherently\/}
many string loops using an operator $F_{a_1, a_2, \cdots, a_n}(\Sigma^{*}_1,
\Sigma^{*}_2, \cdots, \Sigma^{*}_n ,x_0, \{l_P\})$.
Upon gauge transformation
by $g$ at the base point, it changes to  $F_{ga_1 g^{-1}, g a_2 g^{-1} ,
\cdots, g a_n g^{-1}}(\Sigma^{*}_1,
\Sigma^{*}_2, \cdots, \Sigma^{*}_n ,x_0, \{l_P\})$. Note that
for coherent insertion, it is crucial to choose the same base point $x_0$
for all string loops.
Up to now, we have been vague about the choice of $\{l_P\}$.
As we will see below, this is actually quite important.

To introduce classical charges into the
system, we use
the {\em untraced\/} Wilson loop operator
\begin{equation}
U^{(\nu)}(C, x_0)= D^{(\nu)} \left( \prod_{l \in C} U_l \right)
\label{untrace}
\end{equation}
  where $C$ is a closed loop around $x_0$ and $\nu$
is an irreducible representation of the gauge group $G$. Heuristically,
after gauge fixing
all the matrix elements in
$U^{(\nu)}(C, x_0)$ can, in principle, be determined by interfering
charged particles in the
representation $\nu$
that traverse $C$ with those that stay at the base point \cite{AlCoMa}.
Like
$F_a$, the operator $U^{(\nu)}(C, x_0)$ is not
gauge invariant.

It was noted in Ref. \cite{qft} that in a phase with
free $G$ charges, and
in the leading order of
weak coupling perturbation theory,
the operator
\begin{equation}
\lim { F_a(\Sigma^{*}, x_0, \{l_P\}) U^{(\nu)}(C, x_0) \over
\langle F_a(\Sigma^{*}, x_0, \{l_P\}) \rangle
\langle tr U^{(\nu)}(C, x_0)}\rangle
= {1 \over n_{\nu}} D^{\nu} \left( a^{k (\Sigma^{*}, C)} \right)
\label{order}
\end{equation}
 where ${k (\Sigma^{*}, C)}$ is linking number
of the surface $\Sigma^{*}$ and the loop $C$ and the limit that
$\Sigma^{*}$ and $C$ are infinitely large and far away is taken.
This suggests that, once a string loop is calibrated to be of flux $a$
along the paths $\{l_P\}$, a subsequent interference experiment
with a charged
particle travelling around $C$ will give the same
non-Abelian Aharonov-Bohm phase.
However,
higher order terms in the weak coupling expansion may spoil
this result \cite{qft}.
Recall that,
in the definition of $ F_a(\Sigma^{*}, x_0, \{l_P\})$, for each plaquette
$P$, there is a long tail of links $l_P$ that connects it to $x_0$.
Consider,
configurations (in three spacetime
dimensions)
in which a single link on the path that connects $x_0$ to $\Sigma^*$
is excited.
This causes the excitation of
four plaquettes and is suppressed by terms that are {\em independent\/} of the
size of $\Sigma^*$ and $C$ or the separation between them.
We emphasize that the issue of the choice of $\{l_P\}$
is not
merely a mathematical
problem concerning the implementation on the lattice, but has a
{\em physical\/}
basis. The excitation of a single link is due to the traversal of
a virtual vortex-antivortex pair
around a link on $l_P$.(FIG. 1.)
The
inserted flux is conjugated in these configurations and these
higher order corrections render the flux uncertain
up to conjugation. For this reason, these operators were discarded
in Ref. \cite{qft}.

However, in the above discussion,
one has implicitly
assumed that the long tails, $\{l_P\}$, from all the
plaquettes finally merge together and connect to the base point
through a single link that is not on the Wilson loop \cite{private}.
We will see below how a more careful choice of $\{l_P\}$ can
overcome the problems caused by vacuum fluctuations and the operators
defined in Eq. (7) are true order parameters for non-Abelian gauge
theories.
Before we discuss our resolution, let us note another problem.
With this implicit choice of long tails,
these quantum fluctuations also destroy the coherence of
flux between various strings. This is because the calibrated flux of
one string may be conjugated while the elements associated
with others are unaffected.
This relative change in flux is highly physical and
does not go away even when we take the trace of our operator.
Consider the operator
\begin{equation}
 A^{\nu}_{a_1, a_2, \cdots, a_n} ( \Sigma^{*}_1, \Sigma^{*}_2, \cdots,
 \Sigma^{*}_n,x_0,\{l_P\};C) = { F_{a_1, a_2, \cdots, a_n}
( \Sigma^{*}_1, \Sigma^{*}_2, \cdots, \Sigma^{*}_n,x_0,\{l_P\}) W^{\nu}(C)
\over \langle F_{a_1, a_2, \cdots, a_n}
( \Sigma^{*}_1, \Sigma^{*}_2, \cdots, \Sigma^{*}_n,x_0,\{l_P\}) \rangle
\langle
W^{\nu}(C) \rangle} . \label{nstring}
\end{equation}
With our implicit choice of $\{l_P\}$, one finds,
contrary to the claim made in Ref. \cite{qft}, that
\begin{equation}
 \lim \langle A^{\nu}_{a_1, a_2, \cdots, a_n}
( \Sigma^{*}_1, \Sigma^{*}_2, \cdots,
 \Sigma^{*}_n,x_0,\{l_P\};C) \rangle =
\left({1 \over n_{\nu}}\right) \chi^{\nu}(a_1 a_2 \cdots a_n) +
 \text{h.o.t.}
 ,
\label{incoherent}
\end{equation}
where $n_{\nu}$ is the dimension of the representation $\nu$
and the higher order terms (h.o.t.) are non-zero.
Taken at face value, our results seem to suggest
that, because of quantum fluctuations,
the construction of order parameters
for non-Abelian gauge theories is a hopeless enterprise.

Now we present our resolution.
Let us look at the choice of
$\{l_P\}$ more closely.
With an ingenious choice of $\{l_P\}$, can
one prevent vacuum fluctuations from conjugating the inserted flux of
a whole string loop
at a low energetic cost? The answer is a
disappointing no. Since all the tails, $\{l_P\}$,
originates from the base point, quantum fluctuations can always conjugate
the flux of a whole string loop just by flipping all the links
from which the tails leave the base point.
Fortunately, the relevant
question really is: Are there choices of
$\{l_P\}$ by which any energetically
inexpensive vacuum fluctuation which conjugates
the {\em calibrated\/} flux of a whole string loop necessarily also conjugates
the flux {\em measured\/} by the Wilson
loop? The answer is yes.
As emphasized at the beginning of this Letter, any experimental determination
of the non-Abelian
Aharonov-Bohm phase essentially
proceeds in two stages, calibration and
measurement. Unless the two are done in a coordinated manner,
it is understandable that one may be fooled by
quantum fluctuations:
Quantum fluctuations in FIG. 1 conjugate
the calibrated flux without
affecting the measuring apparatus, thus preventing the recovery of
the calibrated flux in the measurement. There are also configurations
in which quantum fluctuations affect the
measuring apparatus but not the calibrating apparatus (FIG. 2).
Note that the configuration shown in FIG. 2
is, in fact, a smooth deformation of that of FIG. 1.
The topology behind these two figures is that the vortex-antivortex
worldline has a non-trivial
linking number with the union of the Wilson loop, the
world sheet of the inserted string, and
$\{l_P\}$.

Having observed this point, quantum fluctuations are
easy to beat. First, note that
if the Wilson loop and the path of calibration were the same, i.e. $
C=l_P P l_P^{-1}$ for some $P$,
the Wilson loop would trivially recover the calibrated element.
The key difficulty
is that, for the operator to be of any use, the tails
${l_P}$'s will inevitably contain links that are not on the untraced Wilson
loop.
Quantum mechanical fluctuations of those links affect the
calibration apparatus without affecting the measurment apparatus
\cite{private}.
The original construction is particularly vulnerable because
there is a single link that belongs to all tails but is not on
the Wilson loop.
Branching clearly helps reduce its vulnerability. Moreover, it may
be a good idea for at least
some of the tails to be initially on our Wilson loop, even though
they must eventually branch out from it.

Consider the configuration shown in FIG. 3 where
the tails are chosen in such a way that many of them beginning from
the base point
are on our Wilson loop initially and branch out one by one
from it.
(This statement requires qualifications: After their
branching out from the Wilson loop,
the tails must not
intersect one another. Also, the Wilson loop must not come close to retrace
itself \cite{Loa} .)
In what follows, we argue
that this construction overcomes all difficulties caused by
quantum fluctuations. In order for quantum fluctuations to
affect the calibration but not the measurement,
links on the Wilson loop must not be excited.
Since by construction the tails that branch out from Wilson loop never
intersect one another after their branching out, to achieve
overall conjugation of the flux of string loops, we must then flip
a link in each tail after it has branched out.
Since the number of tails that branch out becomes large
as
$\Sigma^*$ and $C$ get large, we must
excite a large number of links and pay a huge
energetic cost.
Therefore,
we conclude that, with the choice of $\{l_P\}$ and $C$ in
FIG. 3, there is no energetically inexpensive
way of conjugating
the flux of a whole string loop without affecting
the Wilson loop.

Having specified our choice of tails, we now compute the higher order
terms. It turns out that at weak couplings, the matter action
as well as any configuration in which any two excited links share
a plaquette can be safely ignored. \cite{AlfMar}
By expanding the gauge action
in terms of character functions and using
the orthogonality
relations between the matrix elements of irreducible representations
at weak couplings, we find that
Eq.\ (\ref{order}) is unspoiled by higher order corrections \cite{Loa}.
Hence, we have found a set of powerful order parameters.
In physical terms, by performing the Aharonov-Bohm interference
experiments
with string loops of various flux in $G$
and with our careful choice of $\{l_P\}$ and $C$, we can figure out what the
unbroken group is. (When a gauge group $G$ is broken into $H$,
the elements of $G$ that are not in $H$ are not associated
with isolated cosmic strings, but with strings that are boundaries of domain
walls.
These domain walls are
unstable and will decay via spontaneous nucleation
of string loops \cite{Vilenk}. Consider the insertion of a
string world sheet of flux $a \not\in H$.
Ultimately, the decay of a domain wall bounded by an $a$ string will be
dominated by the string $b$ with the least string tension such that $ab^{-1}
\in H$
and the flux measured by the untraced Wilson
loop will be $ab^{-1}$ rather than $a$ \cite{qft}. We can,
therefore, figure out
what the manifest symmetry group is through a
careful interpretation
of the results of our measurements.)

However, complications arise when the unbroken subgroup, $H$,
is not normal in $G$ because the embedding of $H$ in $G$ now
depends on the value of the Higgs field and has no invariant meaning
of its own. i.e. one should consider $H$ as $H(\phi)$.
One way out is to perform gauge fixing. Suppose
$G$ act transitively
on the Higgs $\phi$. Without loss of generality, one can consider
the gauge fixed insertion operator, $F_a(\Sigma^{*}, x_0, \{l_P\},\phi_0)$,
where the Higgs field $\phi= \phi_0$. When this operator is
inserted in a Green function
with gauge invariant operators, it will have the same effect as the
operator, $ {1 \over |G|} \sum_{g}
F_{gag^{-1}}(\Sigma^{*}, x_0, \{l_P\},R(g)\phi_0)$. For a
particular value of
$\phi$, this reduces to $ {1 \over |H|}
\sum_{h \in H(\phi)}
F_{hah^{-1}}(\Sigma^{*}, x_0, \{l_P\},\phi)$.
This is our second set of order parameters. The details of
our construction will be presented
elsewhere \cite{Loa}.

Finally, we remark that
subtleties in gauge fixing\cite{Bucher,Lee}
also
occur in a Hamiltonian formulation of the quantum mechanics of
non-Abelian
vortices. This subject has shown to be of some conceptual
interest\cite{LoLePr} . In a forthcoming paper \cite{Lo},
we discuss the role of base point in the Hamiltonian
formulation and show that
the quantum mechanics of a single vortex
on a cylinder or a torus is exactly solvable.

We are indebted to J. Preskill for bringing this problem to our attention
and for stimulating discussions. Helpful conversations with M. Bucher,
K.-M. Lee and J. March-Russell are also gratefully acknowledged. This
work was supported in part by DOE DE-FG02-90ER40542.

\begin{figure}
\caption{A worldline of vortex-antivortex pair winds around the tail
$l_P$, leading to the conjugation of the calibrated flux at the
cost of exciting one link. i.e. four plaquettes.}
\label{fig1}
\end{figure}

\begin{figure}
\caption{A deformation of the configuration shown in FIG. 1.
The worldline of the vortex-antivortex now winds around the Wilson loop,
thus affecting the measurement, but not the calibration apparatus.}
\label{fig2}
\end{figure}

\begin{figure}
\caption{With a coordinated choice of the Wilson loop and $\{l_P\}$,
any energetically inexpensive excitations that affect the calibration process
necessarily affect the measurement process and vice versa.}
\label{fig3}
\end{figure}
\end{document}